# Identification and Prediction of Photoplasticity in Semiconductors Using Feature Engineering and Machine learning

Huicong Chen [a,1], Mingqiang Li [a,1], Zheyuan Ji [a,1], Yu Zou [a,*]

[a] *Department of Materials Science and Engineering, University of Toronto, 184 College Street, Toronto, M5S 3E4, Canada*

[1] *These authors contributed equally*

*Corresponding author:* Email: mse.zou@utoronto.ca (Y.Z.)

**ABSTRACT**

Photoplasticity—the light-induced change in plastic deformation—plays a pivotal role in the mechanical durability and manufacturing of semiconductor materials. Yet, its governing mechanisms remain incompletely understood, owing to the interplay of coupled multiphysics factors. Here, we conduct high-throughput nanoindentation measurements to compile a dataset of paired hardness values in dark and light conditions. Then, we engineer physics-informed descriptors spanning electrical, mechanical, and optical properties, and identify the ten most informative features, including bandgap, breakdown field, and refractive index, to enable an interpretable machine-learning framework that yields transferable design rules for light-tunable semiconductor mechanics. By identifying and predicting photoplasticity in semiconductors, this work provides a practical pathway for extracting mechanism-linked, transferable guidelines to engineer light-responsive mechanical behavior in semiconductor materials and devices.

# I. INTRODUCTION

Light-induced changes in mechanical response are becoming increasingly critical, as semiconductors are pushed into regimes where optical, electrical, and mechanical loads act simultaneously in power electronics, detectors, and optomechanical components [1-3]. It has been found that optical excitation reversibly alters yield strength, flow stresses, and hardness in many semiconductors and drives photo-softening or photo-hardening, revealing that non-thermal electronic excitation could couple strongly to the mechanical deformation [4,5]. Recent work has renewed broader interest in functionalizing defects and dislocations to achieve stimulus-controlled mechanical responses, motivating a modern re-examination of photoplasticity using quantitative, operando micromechanical methods and data-centric modeling [6-8].

In crystalline semiconductors, photoplastic responses are governed primarily by how photoexcited carriers interact with dislocations and other defect states. Studies on ZnO quantified reversible light-induced strengthening and mapped strong dependences on light intensity, crystal orientation, and temperature, providing evidence for carrier-mediated mechanisms [9-11]. In III–V systems, light induced dislocation glide in GaAs/AlGaAs heterostructures revealed a sharp excitation threshold and highlighted the role of nonequilibrium carriers in modifying the effective lattice friction on dislocation motion [12]. Additionally, the recombination-enhanced framework established that nonradiative electron–hole recombination at defects locally transfers energy into the lattice and accelerates defect reactions and dislocation transport, while complementary models emphasize charging/screening of dislocation cores and surrounding point defects in II–VI compounds [13-17]. These indicate that the sign and magnitude of photoplasticity depend on the competition between carrier-assisted defect mobility (photo-softening) and carrier/charge-state–driven increases in glide resistance (photo-hardening). Recent nanoindentation and creep-based investigations in ZnS demonstrate particularly large and reversible light sensitivities, and first-principles work links these responses to carrier trapping and reconstruction at dislocation cores that modify Peierls energy barriers [6,18-22]. Nevertheless, the previous studies have underscored the practical challenge: photoplasticity is multi-factorial, and the governing variables have rarely been co-measured.

Up to date, published data span disparate probe volumes, contact geometries, microstructures, light spectra and fluxes, and temperatures, with different subsets of carrier lifetimes, mobilities, defect densities, and baseline mechanical properties [19,22-26]. As a result, it remains unclear which combinations of electronic-structure, carrier-kinetic, defect/dislocation, and mechanical parameters dominate Photoplasticity. The photo-nanoindentation method offers a uniquely efficient route to generate harmonized, light-on/light-off hardness data with controlled contact mechanics, enabling systematic comparison and, crucially, the possibility of learning structure–property–response relations from curated datasets [6,21,22,27,28]. Additionally, machine-learning (ML) methods have begun to reshape mechanical characterization and property prediction, enabling the identification of informative descriptors through feature engineering [29-32].

In this study, we identify and predict photoplasticity in a wide range of semiconductors by integrating photo-nanoindentation with physics-informed feature engineering and machine learning. We assemble a harmonized dataset capturing light-induced changes in hardness and construct descriptors that encode (i) electrical, (ii) mechanical, and (iii) optical effects. We then benchmark state-of-the-art regression models and interpret their predictions using transparent attribution analyses, identifying the dominant descriptors that rationalize light-driven hardening and softening across semiconductor families.

## II. METHODS

### A. Nanoindentation in the light

**Figure 1a** shows the nanoindentation performed in the dark or light conditions. The LEDs were used as the light source, and a modular light unit (280-940 nm) was designed and integrated into the nanoindentation system (iMicro, KLA). The LED wavelengths were selected according to the semiconductor bandgaps (e.g., 365 nm for ZnS with a bandgap of ~3.54 eV), as shown in **Table S1**. Load–depth data were acquired for every indent, and hardness (*H*) was calculated using the Oliver–Pharr method [33]. The photoplasticity was characterized by calculating the relative difference between the hardness tested in the dark and light, i.e., $\Delta H = \frac{H_{light} - H_{dark}}{H_{light}} \times 100\%$. **Figure 1b-d** show representative nanoindentation load–depth curves for ZnS, Si, and GaAs, illustrating light-induced hardening, softening and minimal hardening, respectively. Source data for all semiconductors examined are provided in the **Supplementary Materials**. It shall be noted that llight-induced hardening/softening depends on both the material condition (including doping/defect chemistry) and light parameters. Each condition in **Table S1** was repeated multiple times, with errors kept within an acceptable range (**Figure 2a**), nevertheless, differences from the literature may persist due to variations in sample state, surface preparation, and experimental light/indentation protocols.

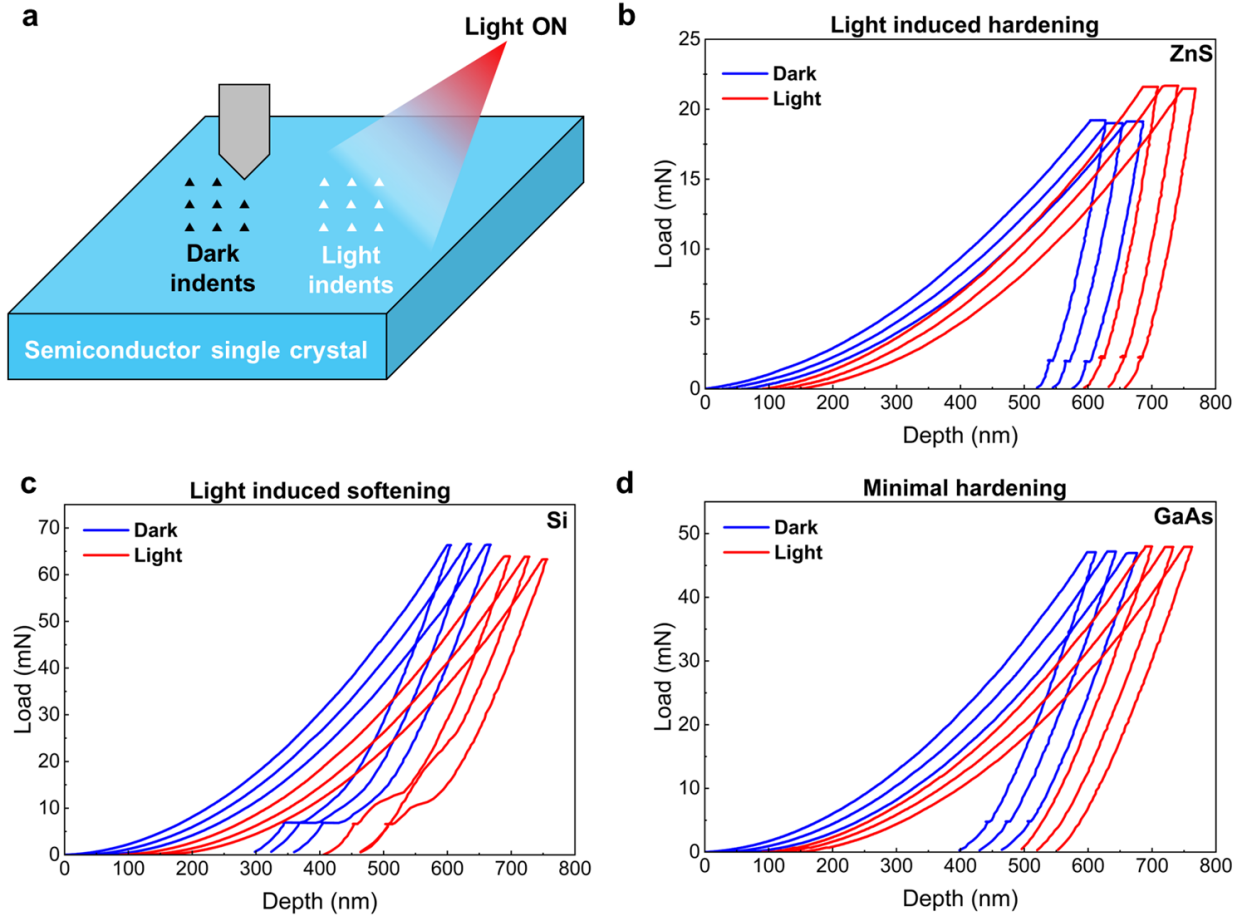

**Figure 1.** (a) Schematic illustration of high-throughput nanoindentations performed in the dark and light conditions. Representative load-depth curves in nanoindentation test of (b) ZnS with light-induced hardening, (c) Si with light-induced softening, and (d) GaAs with minimal hardening, i.e., light insensitive. For each sample, three curves are shown for tests conducted in the dark and light.

## B. Feature engineering

To predict the light-induced hardness change $\Delta H$, we built a physics-informed descriptor pool (**Table S2**), which is organized into three blocks: (i) electrical descriptors (D1–D6) capture the likelihood of photocarrier generation and transport to dislocation cores; (ii) mechanical descriptors (D7–D16) encode baseline stiffness and lattice friction governing dislocation nucleation and glide; (iii) Optical descriptors (D17–D20) represent light–matter coupling, carrier lifetime, and electrostatic screening relevant to charged-defect interactions. As a rapid, model-agnostic screen, we quantify the bivariate association between each physics-informed descriptor $X$ and the light-induced hardness change $Y = \Delta H$ using the Pearson correlation coefficient,

$$\rho_{X,Y} = \frac{cov(X,Y)}{\sigma_X \sigma_Y} = \frac{\sum_{k=1}^{n}(x_k - \bar{x})(y_k - \bar{y})}{\sqrt{\sum_{k=1}^{n}(x_k - \bar{x})^2} \sqrt{\sum_{k=1}^{n}(y_k - \bar{y})^2}} \quad (1)$$

To quantify descriptor contributions beyond bivariate trends, we apply tree-based SHAP (Shapley Additive Explanations) [34] to the trained decision-tree ensemble for $\Delta H$. SHAP represents the model prediction as an additive decomposition:

$$f(x) = \phi_0 + \sum_{i=1}^{M} \phi_i \quad (2)$$

where, $\phi_0 = E[f(X)]$ and $\phi_i$ is the Shapley-consistent contribution of descriptor i. Formally,

$$\phi_i = \sum_{S \subseteq F \setminus \{i\}} \frac{|S|!\,(M - |S| - 1)!}{M!} [v(S \cup \{i\}) - v(S)] \quad (3)$$

with $v(S) = E[f(X)|X_S = x_S]$. Global importance was summarized as $I_i = \frac{1}{n}\sum_{k=1}^{n} |\phi_i^{(k)}|$, while sign-resolved attributions indicate whether increasing a descriptor tends to promote light-induced hardening or softening.

### III. RESULTS

**Figure 2a-b** summarize the distribution of the hardness variation $\Delta H$ with hardness for various semiconductors. The dataset spans -3.96% to 8.47%, indicating that light can increase or decrease the hardness depending on types of semiconductors. Notably, II–VI compounds exhibit the largest light-induced hardening, with ZnS (8.47%), CdS (7.71%), and ZnTe (5.53%) showing pronounced positive $\Delta H$. These strong responses are consistent with recent nanoindentation reports of the photoplasticity in ZnO, ZnS, and CdS, attributed to defect-mediated carrier interactions that suppress dislocation mobility under light [19]. First-principles studies further support this mechanism for ZnS, suggesting that photoexcited carriers can be trapped at charged dislocation cores, driving core reconstructions and increasing the Peierls barrier, i.e., enhancing lattice resistance to glide, consistent with a hardening response. [24]. By contrast, several covalent and III–V systems show only modest changes near zero (e.g., GaAs 0.35%, GaSb 0.50%, InSb 0.81%, InP 0.92%), while Si (-3.96%), Ge (-2.63%), and InAs (-1.15%) exhibit measurable softening, suggesting that in some materials light reduces effective barriers for defect motion or activates alternative deformation pathways rather than increasing lattice friction.

As shown in **Figure 2c-d**, the distribution is centered close to zero with a median around 0.7% and an interquartile range from -0.02% to 3.31%, indicating that modest hardening is common while a small number of II–VI materials form a strong positive tail. This pattern is consistent with the broader view that photoplasticity is amplified when dislocations are electrically active/charged and strongly coupled to carrier redistribution—an effect that can also be tuned by other non-mechanical stimuli such as an electric field, which directly controls dislocation motion in ZnS [6]. Overall, the results confirm that light-induced hardness changes are measurable, sign-variable, and material-dependent. This motivates a central requirement for predictive modeling: descriptors must encode electronic excitation and carrier transport, electrostatic screening and defect interactions, and baseline glide resistance, rather than treating light as a purely categorical perturbation [19].

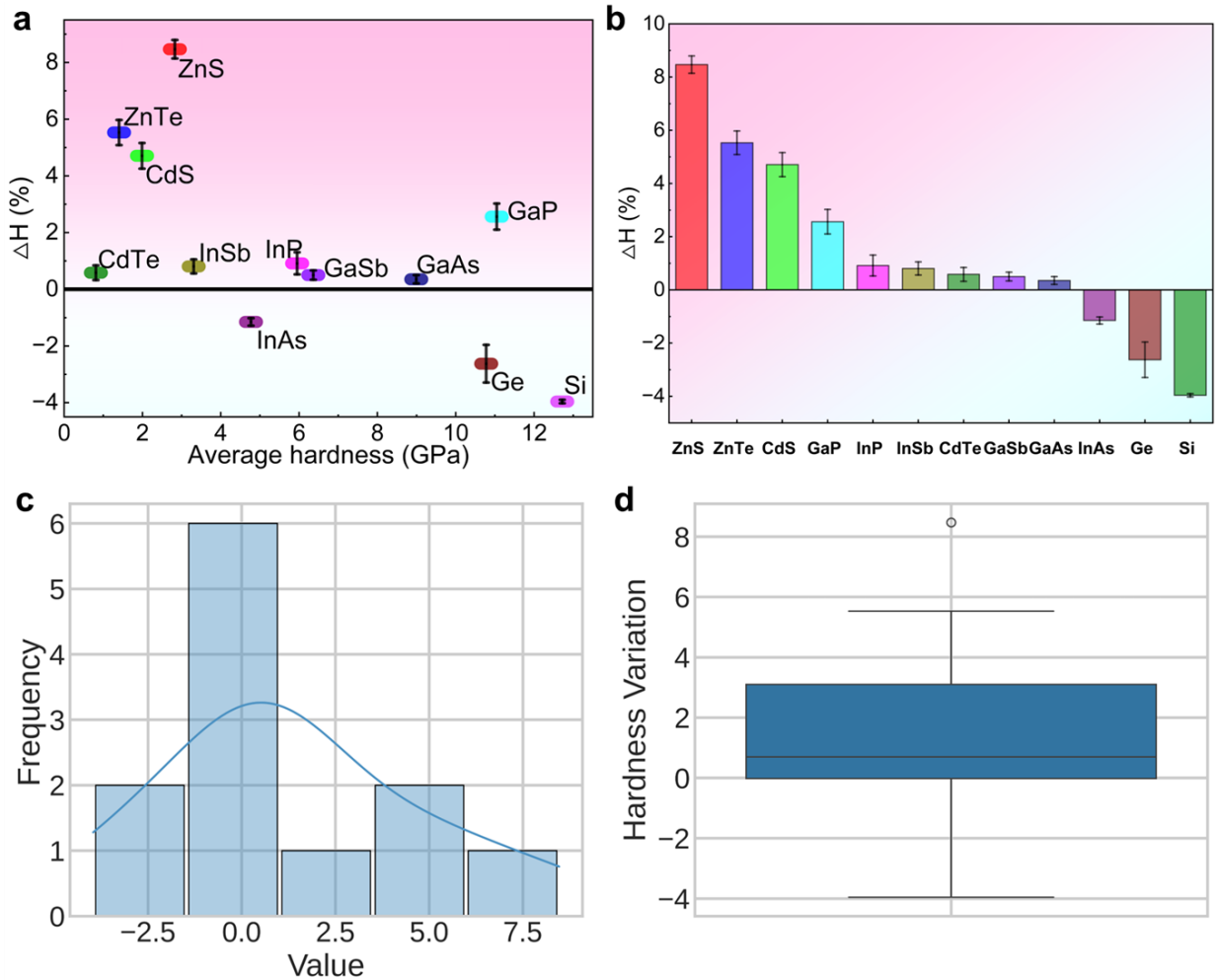

Figure 2 The change of hardness (ΔH) versus (a) average hardness measured under light and (b) types of semiconductors in this study various semiconductors. (c) Histogram of the target ΔH, illustrating its overall distribution across the dataset and the median and (d) interquartile range, providing a compact view of central tendency, spread, and extreme values.

**Figure 3** identifies the descriptors most strongly associated with $\Delta H$. Bandgap (D1) and breakdown field (D2) show the largest positive correlations (r ≈ 0.83 and 0.82), while infrared refractive index (D18), shear coefficient (D8), bond-bending force constant (D13), hole mobility (D5), and dielectric constant (D20) are among the most negative correlates (| r |≈ 0.64–0.77). The substantial inter-descriptor collinearity (e.g., D1–D2) suggests that correlation magnitudes reflect clustered material-property trends and should be interpreted in a multivariate context rather than as single-variable drivers.

**Figure 4** complements the correlation analysis by providing a tree-based SHAP ranking that captures nonlinear effects and allocates importance across correlated predictors. In contrast to the Pearson screening, the model assigns the highest global importance to dielectric constant (D20) and infrared refractive index (D18), followed by carrier-transport descriptors (electron mobility D4, saturation drift velocity D3, and hole mobility D5), and then bandgap (D1); secondary contributions arise from D13, D6, D15, and D14. This ordering suggests that excitation-related descriptors (D1–D2) primarily set enabling conditions for photoexcitation, whereas screening/polarizability and carrier kinetics more directly regulate the magnitude and sign of ΔH after accounting for multicollinearity. Consistent with Figure 3, higher D20 and D18 generally correspond to more negative SHAP contributions (lower predicted $\Delta H$), implying that stronger dielectric screening attenuates light-induced perturbations of dislocation electrostatics.

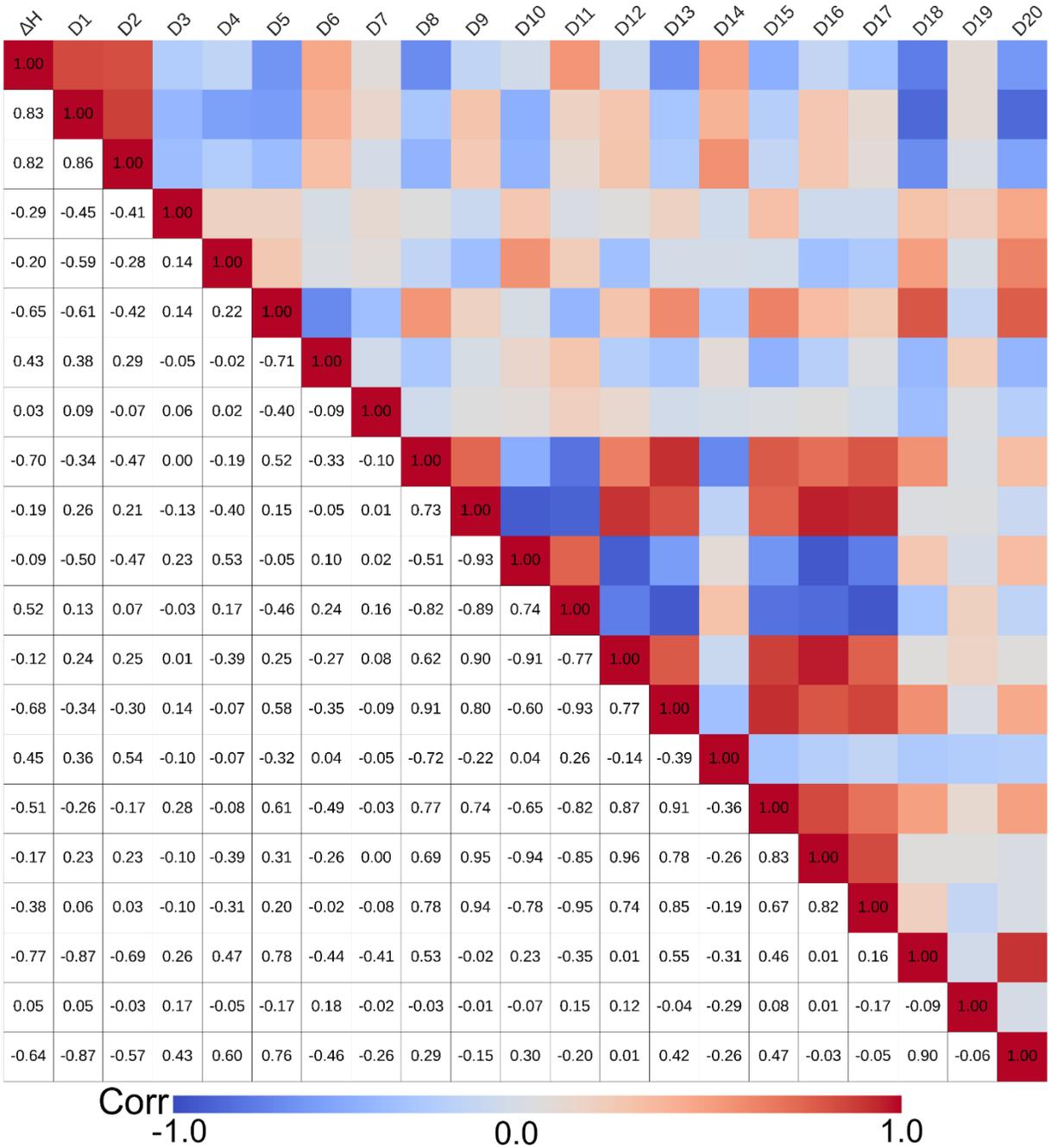

**Figure 3**. Heatmap of the Pearson correlation coefficient matrix among the original descriptors, providing a quick visual overview of how strongly the descriptors relate to one another, revealing patterns of similarity and shared behavior across the feature set.

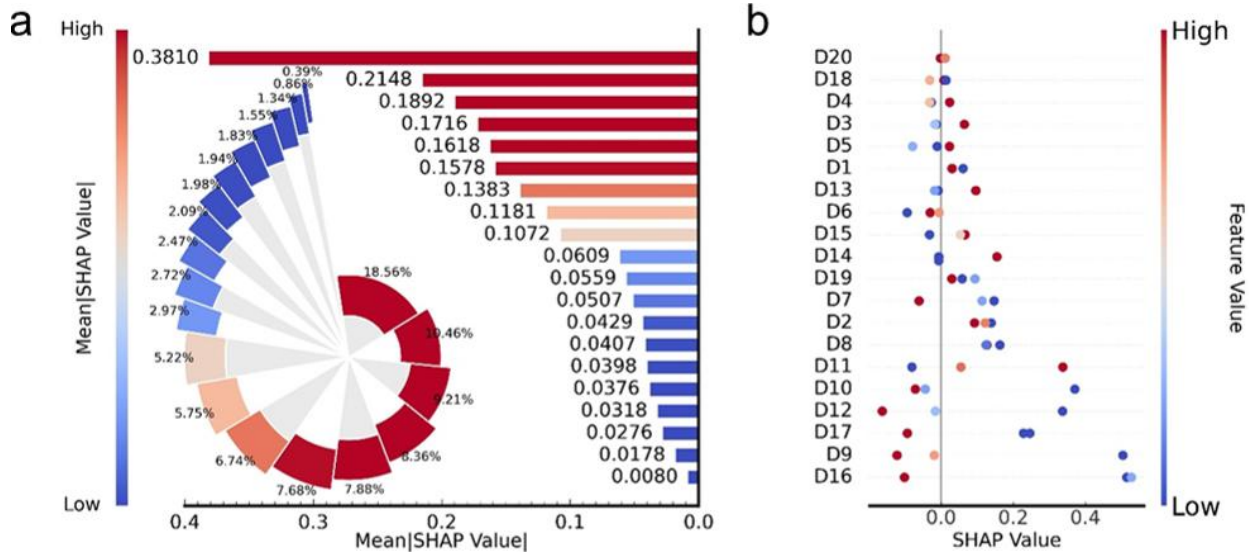

**Figure 4.** SHAP-based feature importance analysis for predicting light-induced changes in hardness. (a) The mean absolute SHAP values rank descriptors by their overall contribution to model output. (b) The SHAP summary plot illustrates the distribution and direction of each descriptor's impact across samples; point color denotes the corresponding feature value (from low to high).

**Figure 5** visualizes the top ten-ranked features (by | r |) and clarifies how they map onto the mechanistic axes proposed above. The dominance of bandgap (D1) is consistent with the widely reported spectral sensitivity of photoplastic phenomena: When light produces a substantial nonequilibrium carrier population, carrier trapping and redistribution near defects/dislocations can strongly perturb the local stress fields and glide barriers [12]. In wide-band-gap II–VI semiconductors, nanoindentation studies have reported giant and reversible photoplasticity under moderate light, with the response mediated by point defects and photocarrier–defect coupling [19,21,22]. In this context, bandgap serves as an effective descriptor for the propensity to sustain large photocarrier perturbations under comparable light conditions, which is consistent with the positive correlation between D1 and Δ$H$ observed here [1,19,21,22]. Breakdown field (D2) ranks comparably to D1, and its interpretive value likely lies in its role as a proxy for the "wide-gap/strongly bonded/low-intrinsic-carrier" regime in which photoexcitation and defect charging can produce a relatively large fractional change in carrier populations and internal electrostatics [1,19,35]. Importantly, this interpretation is reinforced by direct demonstrations that dislocation motion in ZnS can be controlled by a non-mechanical stimulus through the charge characteristics of dislocation cores [6]. Specifically, experimental observations show that an applied electric field

can drive dislocation motion in ZnS, with the effect attributed to non-stoichiometric, charged dislocation cores and field-dependent glide barriers [6,35]. This result is central here because it supports a unified picture: if dislocation mobility is sensitive to charge state and local electrostatics, then descriptors tied to carrier generation and field tolerance (D1–D2) should correlate with the likelihood and magnitude of light-induced hardening or softening [6,35].

Among the negatively correlated features, dielectric constant (D20) and infrared refractive index (D18) are particularly informative because they encode polarizability and screening. [35]. Many photoplasticity models invoke charged dislocations and charged point defects; in such frameworks, the extent to which photocarriers and bound charges modify dislocation–defect interactions depends on screening [13,35,36]. The strong negative association of $\Delta H$ with D20 and D18 is therefore consistent with a mechanism where large hardening is favored when electrostatic interactions around charged dislocations are less effectively screened, enabling light-driven carrier redistribution to more strongly perturb dislocation mobility [6,35]. This is qualitatively aligned with first-principles analyses in ZnS that explicitly treat the role of photoexcited carriers in modifying dislocation-core structure and charge-dependent glide barriers, including changes in Peierls-related resistance to motion [18,20,37-39]. The ranking of hole mobility (D5) provides a complementary handle on carrier kinetics. Mobility governs how rapidly photocarriers can reach and interact with dislocations and defect states during indentation, and it can therefore bias the competition between light-induced hardening (e.g., via carrier trapping at charged cores and increased lattice friction) and light-induced softening (e.g., via carrier-assisted defect motion) [13,36]. Recent mechanistic discussions of photoplasticity in II–VI semiconductors explicitly emphasize the role of carrier populations, including holes, and their coupling to dislocation glide barriers and bond energetics, supporting the physical relevance of mobility descriptors. In addition, experimental reports of spectrally dependent positive and negative photoplastic responses have been interpreted in terms of recombination- or radiation-enhanced dislocation processes, highlighting that transport/recombination pathways can flip the sign of $V_H$ depending on the dominant electronic transition channels [12,13,36].

The emergence of shear coefficient (D8) and bond-bending force constant (D13) indicates that baseline resistance to shear and bond-angle distortion conditions how strongly light-coupled electronic effects manifest in indentation hardness [18,19,21,22]. In practice, these variables are also entangled with bonding class: covalent semiconductors tend to exhibit high bond-angle stiffness and shear resistance, yet in our dataset include the cases of measurable softening. Thus, D8 and D13 likely act as deformation-mode selectors—capturing the underlying ease of dislocation glide under indentation—rather than directly encoding the electronic perturbation itself [20,37-39]. Their importance is consistent with the broader studies on the II–VI semiconductors that optical excitation perturbs plasticity primarily through its impact on dislocation processes, which are ultimately expressed against a mechanical baseline set by lattice resistance and bonding [1,18,19,21,22,35]. Additionally, bond energy (D11) and dislocation core bonding (D15) suggest that bond robustness and core stability are non-negligible descriptors for light-tunable hardness [18,20,21,37,38]. Recent work argues that increased carrier concentrations can enhance effective bond energetics and elevate barriers for dislocation glide in II–VI systems, which is consistent with the positive association between $\Delta H$ and D11 and with the broader hardening trends observed for ZnS, CdS and ZnTe [18,19,21,22]. At the same time, **Figure 3** indicates that bonding descriptors are correlated with several other mechanical and optical parameters; therefore, their Pearson ranking should be interpreted as evidence that the bonding–screening–transport manifold is relevant, not that any single scalar measure of bonding uniquely determines $\Delta H$ [18,20,37-39].

**Figure 3** and **Figure 4** motivate two methodological choices that are carried forward in the machine-learning section. First, the top-ranked features naturally organize into three mechanistic groups—excitation propensity (D1–D2), screening/transport (D5, D18, and D20), and lattice/bonding resistance (D8, D11, D13, and D15)—which mirrors the physical decomposition used for feature engineering [18,19,21,22,35]. Second, because **Figure 3** shows pronounced multicollinearity, a predictive model must be robust to correlated inputs and capable of capturing interactions (e.g., excitation × screening), rather than relying on univariate trends [18,21,22]. This is particularly important given that recent top-tier experiments directly demonstrate that dislocation motion in ZnS can be manipulated by external fields through charged-core physics, implying that predictive descriptors must encode coupled electronic–defect–mechanical effects rather than treating light as a purely categorical perturbation [6,40,41].

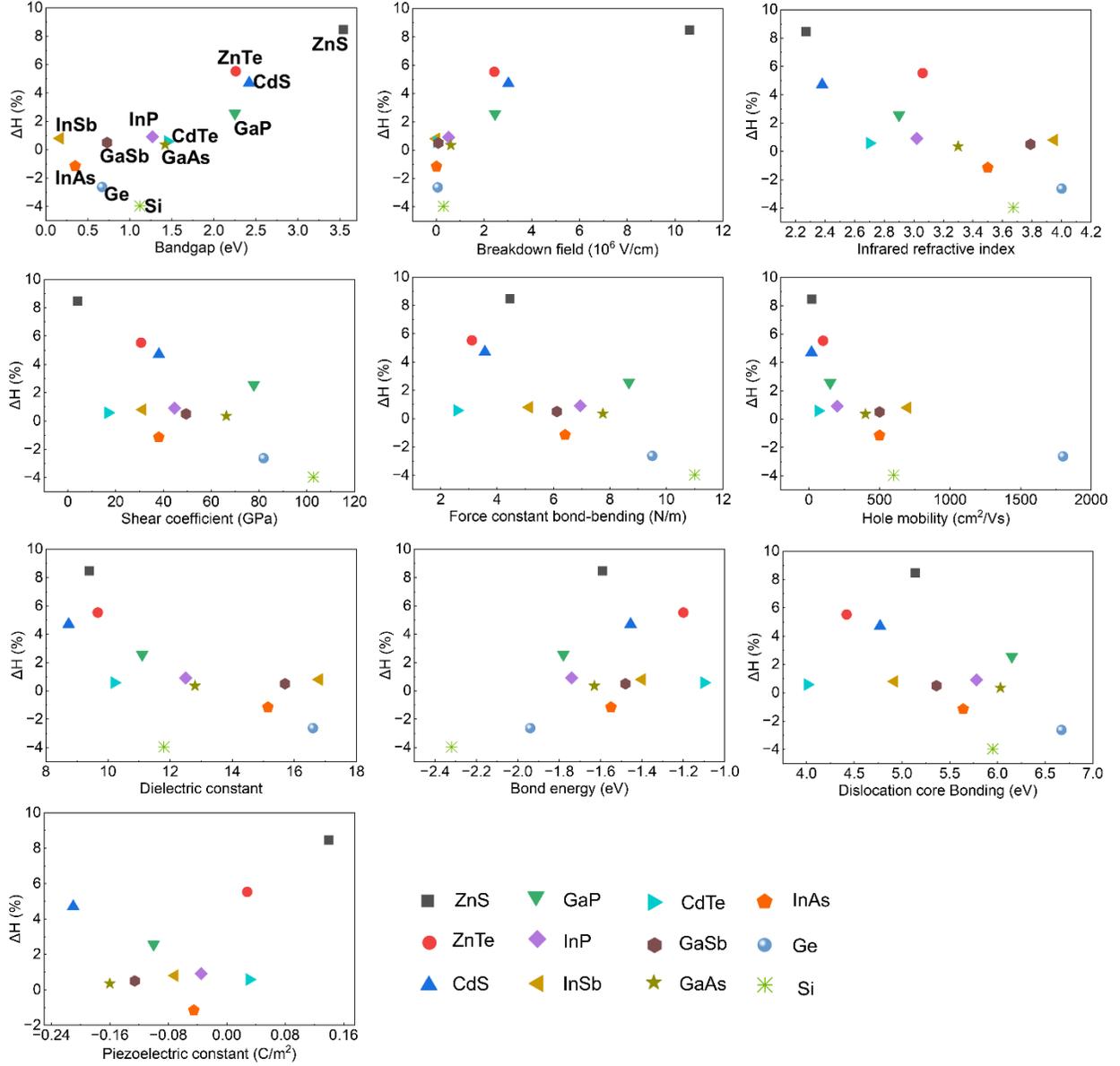

**Figure 5.** Top ten features and their relationships to hardness variations (Δ*H*), showing the effects of key electrical, mechanical, and optical descriptors on the magnitudes of the light-induced hardening or softening effects.

## IV. CONCLUDING REMARKS

In summary, this work establishes a data-driven framework to characterize and rationalize photoplasticity in a wide range of semiconductors using harmonized photo-nanoindentation data and physics-informed descriptors. The compiled dataset shows that light-induced hardness changes are sign-variable, and strongly material dependent, spanning from -3.96% to 8.47%, with a pronounced hardening in II–VI compounds. Feature analysis identifies a consistent set of controlling factors that align with prevailing mechanisms: excitation propensity (i.e., bandgap and breakdown field), carrier transport and electrostatic screening (i.e., mobility, dielectric constant, and refractive index), and baseline bonding strength and lattice friction for dislocation motion ((i.e., bond energy and dislocation core bonding) that governs dislocation nucleation and glide. This work provides a practical route to extract transferable, mechanism-linked design rules for engineering light-tunable manufacturing processes and mechanical durability in a wide range of semiconductor materials and devices.

## ACKNOWLEDGMENTS

The authors acknowledge Natural Sciences and Engineering Research Council of Canada Discovery (Grant No. RGPIN-2018–05731), NSERC Alliance International Collaboration grant (ALLRP 598774–24), Digital Research Alliance of Canada (RRG #5200), the Canadian Foundation for Innovation, John R. Evans Leaders Fund (JELF) 38044, and Data Sciences Institute Catalyst Grant from the University of Toronto.

## DATA AVAILABILITY

The data are not publicly available. The data are available from the authors upon reasonable request.